\documentclass[a4paper,11pt]{article}
\pdfoutput=1 

\usepackage{jinstpub} 
\title{Effects of the electronic threshold on the performance of the RPC system of the CMS experiment}

\author[n,1]{W. Elmetenawee \note{Corresponding author.}}
\author[a]{,A. Samalan}
\author[a]{,M. Tytgat}
\author[a]{,N. Zaganidis}
\author[b]{,G.A. Alves}
\author[b]{,F. Marujo}
\author[c]{,F. Torres Da Silva De Araujo}
\author[c]{,E.M. Da Costa}
\author[c]{,D. De Jesus Damiao}
\author[c]{,H. Nogima}
% ...

\affiliation[a]{Ghent University, Dept. of Physics and Astronomy, Proeftuinstraat 86, B-9000 Ghent, Belgium}
\affiliation[b]{Centro Brasileiro Pesquisas Fisicas, R. Dr. Xavier Sigaud, 150 - Urca, Rio de Janeiro - RJ, 22290-180, Brazil}
\affiliation[c]{Dep. de Fisica Nuclear e Altas Energias, Instituto de Fisica, Universidade do Estado do Rio de Janeiro, Rua Sao Francisco Xavier, 524, BR - Rio de Janeiro 20559-900, RJ, Brazil}
% ...
\affiliation[n]{INFN, Sezione di Bari, Via Orabona 4, IT-70126 Bari, Italy.}
% ...

\emailAdd{walaa.elmetenawee@ba.infn.it}

\abstract{Resistive Plate Chambers have a very important role for muon triggering both in the barrel and in the endcap regions of the CMS experiment at the Large Hadron Collider (LHC). In order to optimize their performance, it is of primary importance to tune the electronic threshold of the front-end boards reading the signals from these detectors. In this paper we present the results of a study aimed to evaluate the effects on the RPC efficiency, cluster size and detector intrinsic noise rate, of variations of the electronics threshold voltage.}


\collaboration[c]{on behalf of the CMS collaboration}

\proceeding{XV Workshop on Resistive Plate Chambers and Related Detectors RPC2020\\
 10-14 February, 2020\\
  Rome, Italy}

\begin{document}
\maketitle
\flushbottom

\section{Introduction}
\label{sec:intro}

Resistive Plate Chambers (RPCs in the following) constitute an important part of the CMS detector at the LHC. Thanks to their good spatial resolution, joined together with an excellent time resolution, they provide a trigger capable of identifying muon tracks both in the barrel and endcap regions of the apparatus~\cite{a,b}. Of course, accurate detector calibrations are crucial to achieve optimal performance during data taking.

In CMS, each endcap RPC chamber is subdivided into three $\eta$ partitions, also called rolls, while barrel RPC chambers are subdivided in two or sometimes in three $\eta$ partitions ($\eta$ being the pseudo-rapidity). Signals coming from each $\eta$ partition are read-out by means of a front-end electronic board, which allows only signals larger than a certain FE discriminator value to be transmitted to the trigger logic. Therefore, fine tuning of these electronic thresholds affects some of the main detector characteristics, like efficiency, cluster size and detector intrinsic noise rate. 

The CMS RPC system has been operating for several years with electronic thresholds which were defined at the time of the first commissioning. These are usually called "default thresholds", differ for each roll and front-end electronics board considered, and are stored in the configuration database. They will be henceforth indicated with $V_{thr-def}$. $V_{thr-def}$ values usually lie in the [205-240] mV range, where 1 mV corresponds to a signal of about 3.2 fC before pre-amplification~\cite{c}.

In order to investigate the possibility to optimize the electronic thresholds, a dedicated scan has been performed during a data taking campaign with cosmic rays. In the scan, the electronic thresholds $V_{thr-app}$ actually applied to the front-end boards were changed, starting from ($V_{thr-def}$ - 5 mV) up to ($V_{thr-def}$ +15 mV), in five steps of 5 mV each. In this paper results about efficiency (computed using the percentage of times a hit on the RPC is present in coincidence with an extrapolated tracks from Cathode Strip Chambers or Drift Tubes, also included in the CMS muon system), cluster size (number of adjacent, fired RPC strips)~\cite{d}, and detector intrinsic noise (counting rate, normalized to the detector surface, measured in a time window opened at random and not in coincidence with the trigger window)~\cite{e}, obtained during this threshold scan, are reported.

\section{Effect of changing front-end discrimination thresholds on RPC detector performance}
\subsection{Effect on RPC efficiency}

The average barrel RPC efficiency measured during the threshold scan is summarized in Table 1, for various values of $V_{thr-app}$. Due to rather less amount of data in the endcap chambers, a detailed study was possible in the barrel region only. As expected, efficiency decreases as $V_{thr-app}$ increases, of about 3.5\% in the threshold range considered.

\begin{table}[h]
\caption{RPC barrel average efficiency, for different applied electronic thresholds.}
\begin{center}
\begin{tabular}{|c|c|c|c|c|c|}
\hline
V$_{thr-app}$ - V$_{thr-def}$ & -5 mV & 0 mV & +5 mV & +10 mV  & +15 mV  \\
\hline
Average Efficiency &  94.5 \% & 93.6 \% & 93.1 \% & 92.2 \% & 91.0 \% \\ 
\hline
\end{tabular}
\end{center}
\end{table}

To exemplify this behaviour, RPC efficiencies of one roll in the barrel (left) and one in the endcap region (right), as a function of the difference between the applied and the default thresholds ($V_{thr-app}$ - $V_{thr-def}$), are shown in Figure 1.

\begin{figure}[h]
\includegraphics[width=7cm]{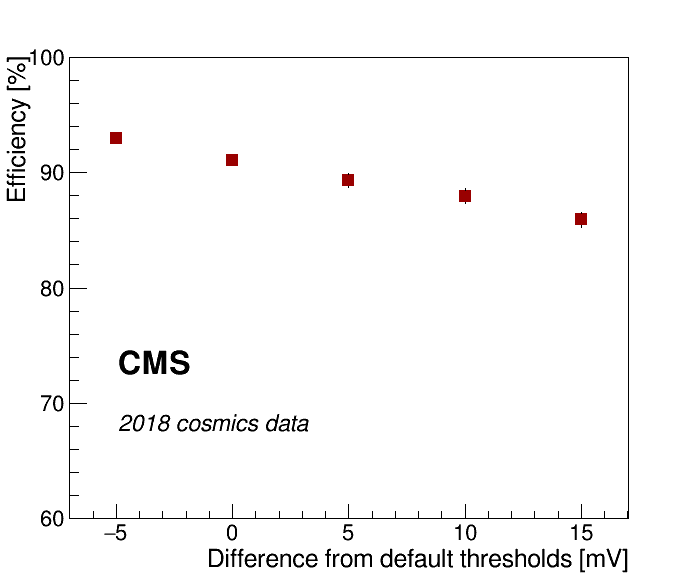}
\qquad
\includegraphics[width=7cm]{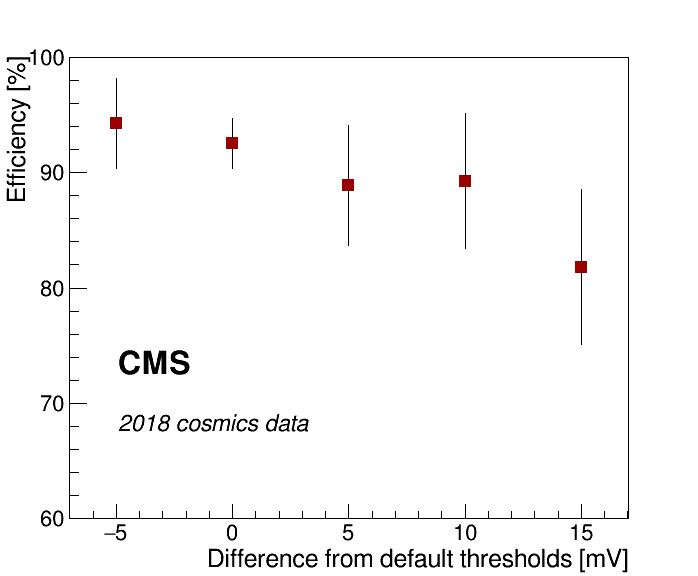}
\caption{RPC efficiency as a function of the difference between V$_{thr-app}$ and V$_{thr-def}$. This figure refers to one roll for the barrel (left) and one in the endcap region (right). Error bars for the endcap roll are larger due to less amount of data.}
\end{figure}

Barrel RPC roll efficiency distributions, measured at the different $V_{thr-app}$ used during the voltage scan, are shown in Figure 2 (left). Efficiency distributions present larger tails to the left of the peak as rolls’ efficiency decrease with increasing the front-end discrimination thresholds. The RPC efficiency variation distribution between the efficiency values measured at $V_{thr-def}$, and the ones obtained with $V_{thr-app}$ = ($V_{thr-def}$ - 5 mV) is shown in Figure 2 (right). Applying a threshold $V_{thr-app}$ 5 mV lower with respect to $V_{thr-def}$ results, in the barrel, in an average efficiency gain of about 0.9\%. The tails in the distributions, containing few entries characterized by high efficiency variations, are caused by less amount of data.

\begin{figure}[h]
\includegraphics[width=7cm]{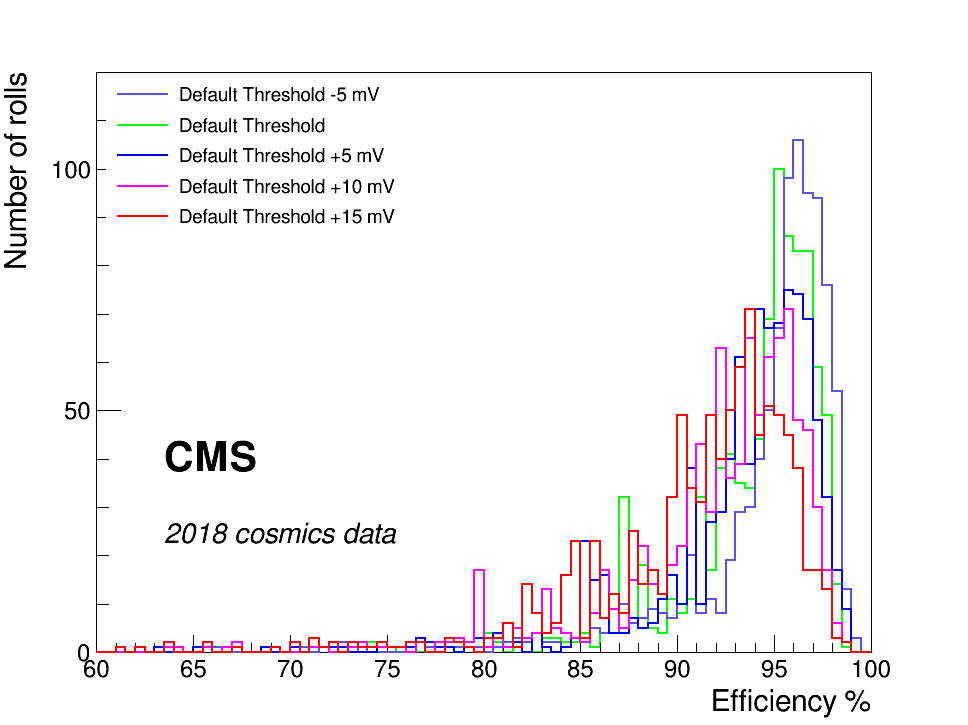}
\qquad
\includegraphics[width=7cm]{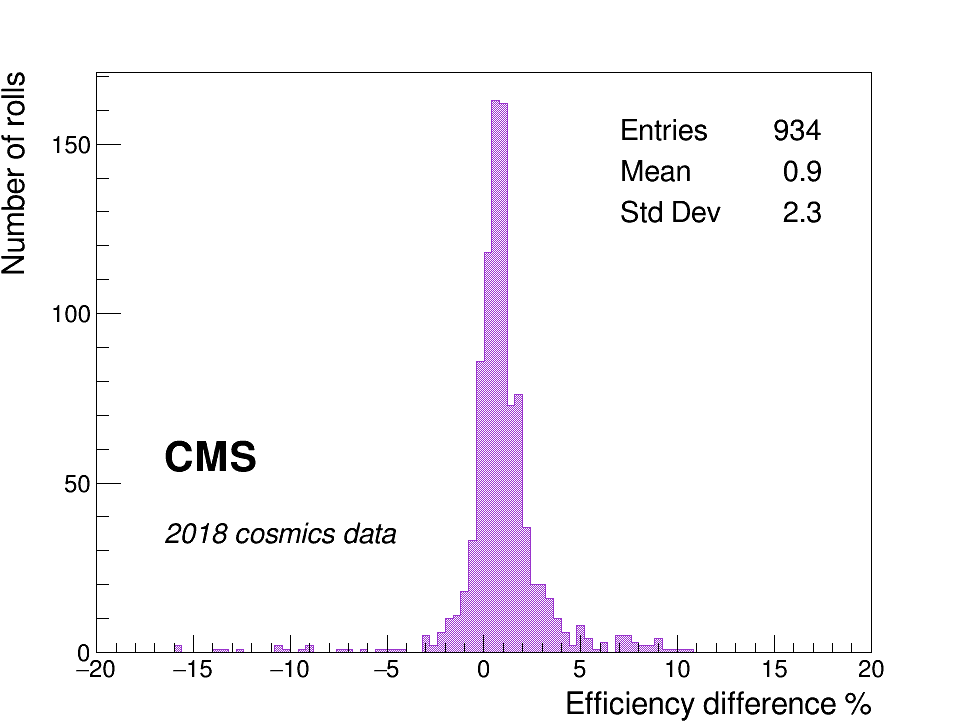}
\caption{Left: Barrel RPC efficiency distributions, measured at the various V$_{thr-app}$ used during the threshold scan. Right: barrel RPC efficiency variation, measured at V$_{thr-app}$ = (V$_{thr-def}$ - 5 mV), with respect to the one obtained with V$_{thr-def}$.}
\end{figure}

\subsection{Effect on RPC cluster size}

The mean values of the barrel RPC cluster size measured applying different values of $V_{thr-app}$ used during the threshold scan are summarized in Table 2. As expected, cluster size decreases as $V_{thr-app}$ increases.

\begin{table}[h]
\caption{Average barrel RPC cluster size, for different values of the applied threshold $V_{thr-app}$}.
\begin{center}
\begin{tabular}{|c|c|c|c|c|c|}
\hline
V$_{thr-app}$ - V$_{thr-def}$ & -5 mV & 0 mV & +5 mV & +10 mV  & +15 mV  \\
\hline
Average cluster size (\# of strips) &  1.76  & 1.70  & 1.67  & 1.62  & 1.58  \\ 
\hline
\end{tabular}
\end{center}
\end{table}

The RPC cluster size of one roll in the barrel and one in the endcap region, as a function of $V_{thr-app}$ is shown in Figure 3. Again, this exemplifies what happens to the cluster size when increasing the electronic threshold in the whole system.

\begin{figure}[h]
\includegraphics[width=7cm]{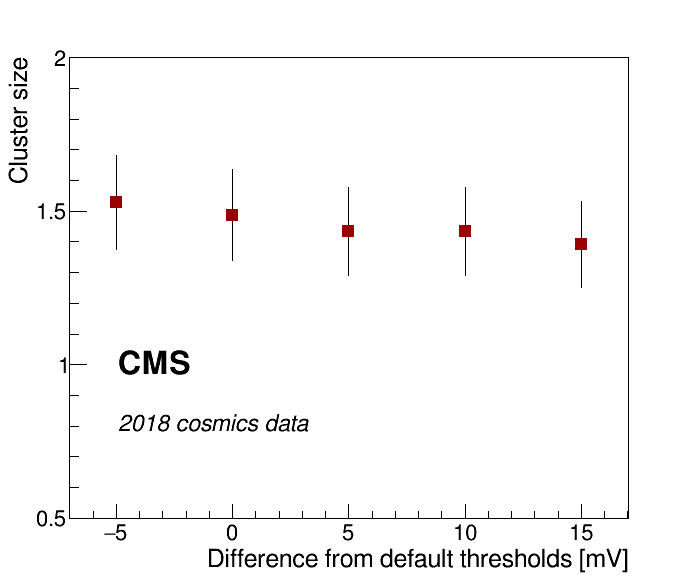}
\qquad
\includegraphics[width=7cm]{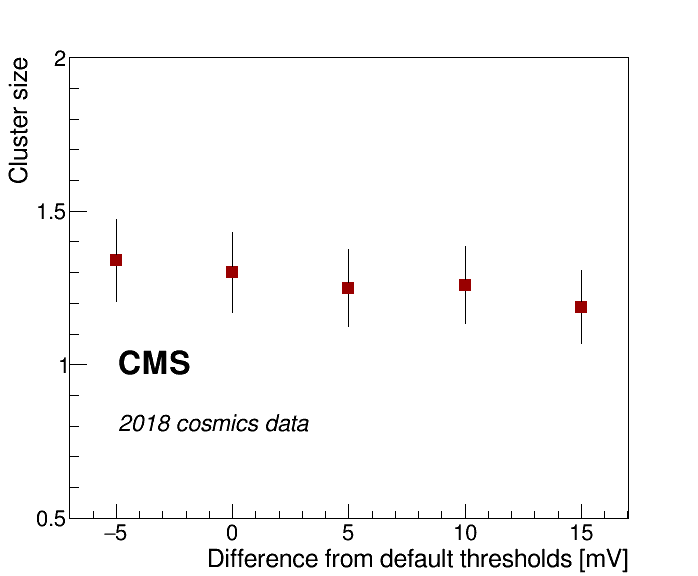}
\caption{RPC Cluster size as a function of the difference between V$_{thr-app}$ and V$_{thr-def}$. Results for one barrel (left) and one endcap rolls (right) are shown.}
\end{figure}

The RPC cluster size distributions, measured at the different $V_{thr-app}$ used during the threshold scan, are shown in Figure 4 (left). Applying a threshold discrimination lower by 5 mV with respect to $V_{thr-def}$ results in a slight increase of the cluster size, quantified in about 0.07 strips for the barrel. This is shown in Figure 4 (right).

\begin{figure}[h]
\includegraphics[width=7cm]{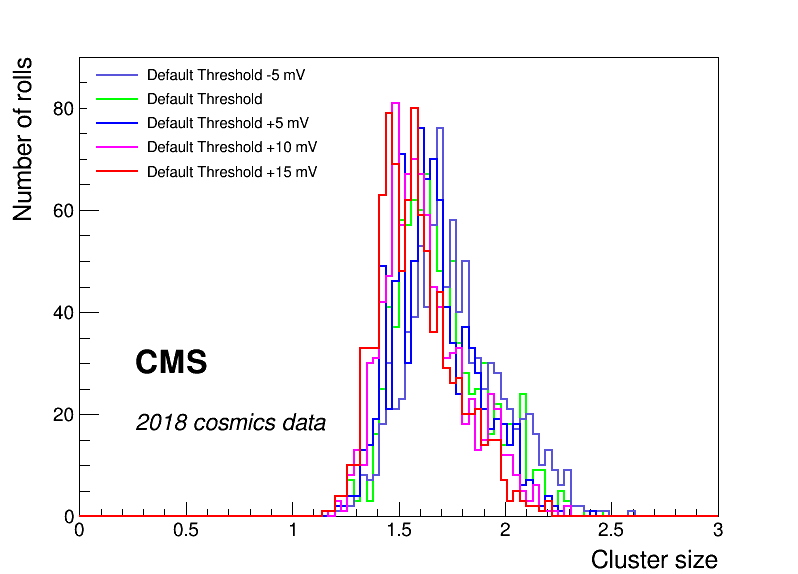}
\qquad
\includegraphics[width=7cm]{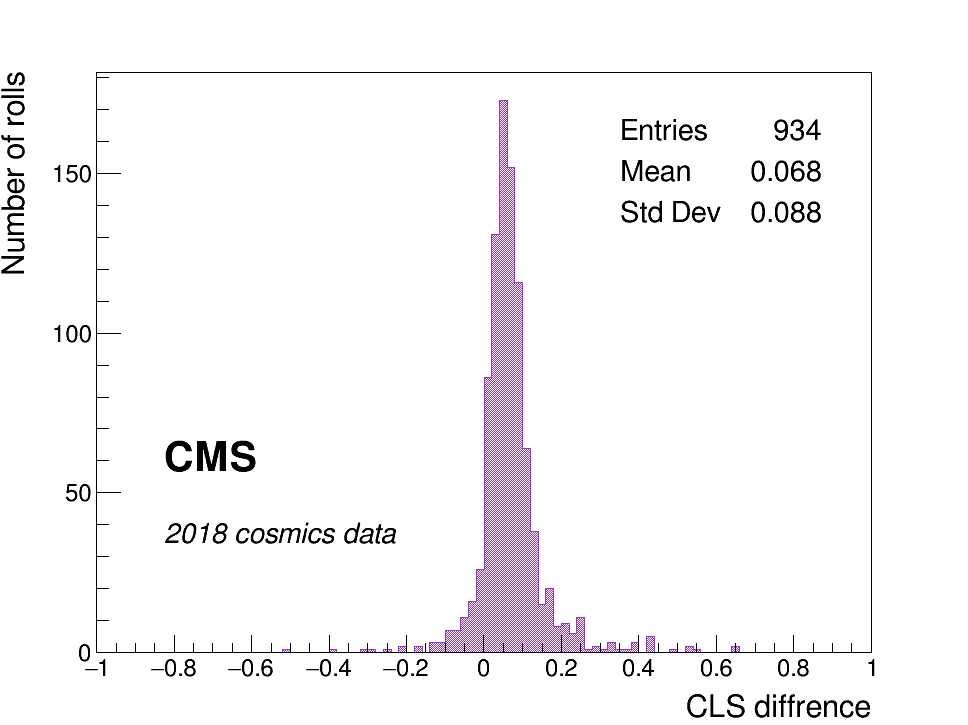}
\caption{(Left) RPC cluster size  distributions measured at the different applied thresholds $V_{thr-app}$ used during the threshold scan. (Right) Variation of the barrel RPC cluster size when decreasing the applied threshold voltage by 5 mV with respect to $V_{thr-def}$.}
\end{figure}

\subsection{Effect on RPC intrinsic noise rate}

The intrinsic RPC noise rate as a function of (V$_{thr-app}$- V$_{thr-def}$) is shown in Figure 5 for one roll in the barrel and one in the endcap region. A quadratic polynomial function is used for the fit. 

The distribution of the difference of the intrinsic noise rate measured for the RPC chambers in the barrel when applying an electronic threshold 5 mV lower with respect to V$_{thr-def}$ is shown in Figure 6. On average, applying 5 mV lower thresholds with respect to V$_{thr-def}$ results in an intrinsic noise rate increase of about 0.04 Hz/cm$^{2}$, which represents around 10$\%$ increase with noise rate measured at the default threshold. 

\begin{figure}[h]
\includegraphics[width=7cm]{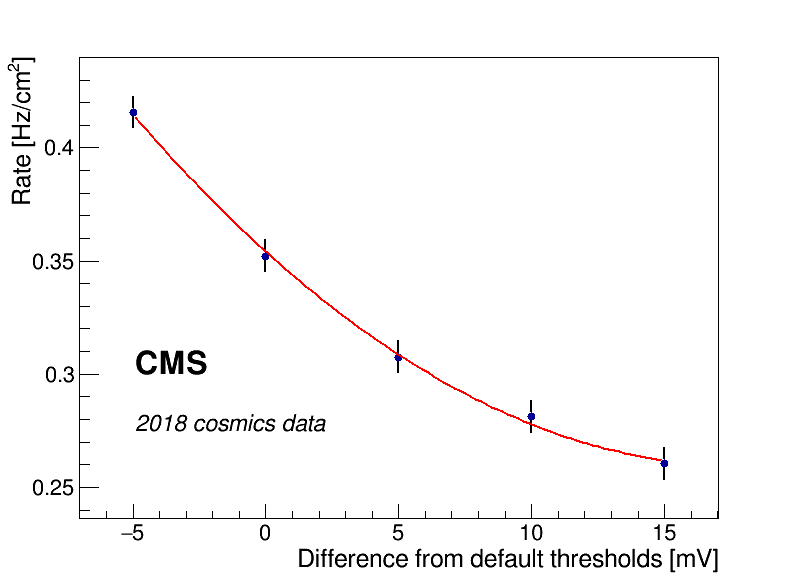}
\qquad
\includegraphics[width=7cm]{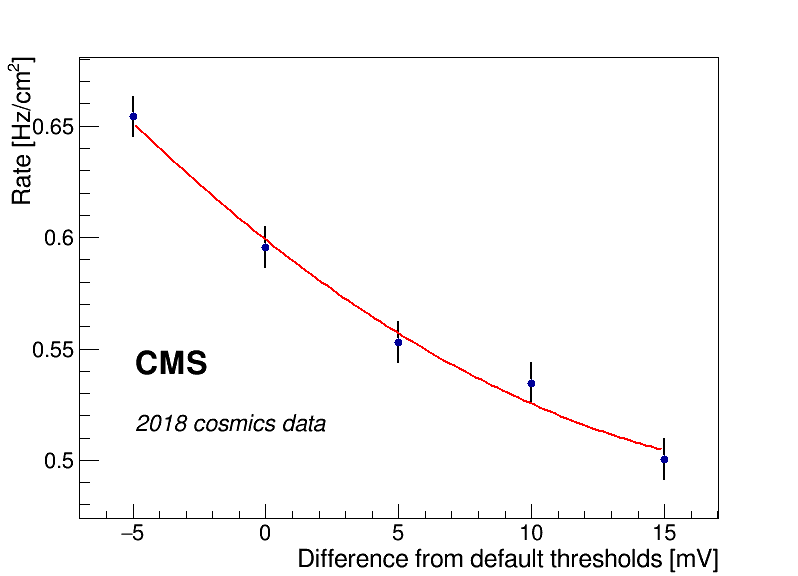}
\caption{RPC intrinsic noise rate as a function of the difference between V$_{thr-app}$ and V$_{thr-def}$. One roll for the barrel (left) and one for the endcap region (right) are shown. In both cases, a quadratic polynomial function is used for the fit.}
\end{figure}

\begin{figure}[h]
\centering
\includegraphics[width=7cm]{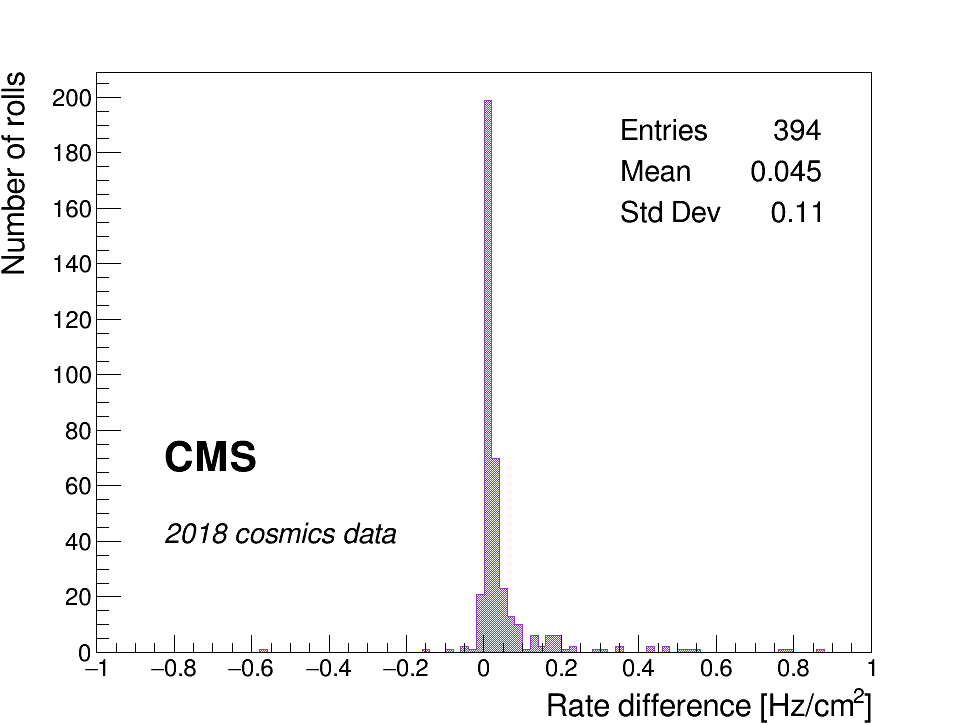}
\qquad
\caption{Variation of the intrinsic barrel RPC rate, measured when applying V$_{thr-def}$ and (V$_{thr-def}$ -5 mV), which results in an intrinsic noise rate increase of about 0.04 Hz/cm$^{2}$.}
\end{figure}

\section{Conclusions}

The results presented here and obtained during the 2018 threshold scan of the CMS RPC system are in agreement with the expectations: a decrease in the applied discrimination thresholds increases the RPC efficiency, which is a benefit for the system but, at the same time, induces an increase in cluster size and intrinsic noise, which could deteriorate spatial resolution and accidentals, respectively. The opposite happens when the threshold is increased. Nevertheless, the $\approx 0.9$\% gain in efficiency consequent to a 5 mV decrease of the threshold with respect to the default one, is to be preferred w.r.t. the corresponding very small increase in cluster size, of around 0.07 strips, and intrinsic noise, roughly 0.04 Hz/cm$^{2}$. This would induce to considerate thoughtfully to adopt such a solution when the CMS RPC system will be restarted after the present shutdown.


\begin{thebibliography}{99}

\bibitem{a}
S. Chatrchyan, et al., \emph{The CMS experiment at the CERN LHC}, JINST3 (2008).

\bibitem{b}
CMS collaboration, \emph{Performance of the CMS muon detector and muon reconstruction with proton-proton collisions at  $\sqrt(s) = 13 TeV$},
2018 JINST 13 P06015 [arXiv:1804.04528].

\bibitem{c}
M. Abbrescia et al., \emph{New developments on front-end electronics for the CMS Resistive Plate Chambers}, \emph{Nucl. Instr. and Meth. in Phys. Res.} {\bf 456} (2000) Pages 143-149.

\bibitem{d}
M.I. Pedraza-Morales et. al., \emph{First results of CMS RPC performance at 13 TeV}, 2016 JINST 11 C12003.

\bibitem{e}
S. Costantini et. al., \emph{Radiation background with the CMS RPCs at the
LHC}, 2015 JINST 10 C05031.

\end{thebibliography}
\end{document}